\documentclass[11pt]{article}
\def\s{\sigma}

\def\r{\ref}
\def\p{\partial}
\def\no{\nonumber}
\begin{document}

\title{{\bf{\Large  String non(anti)commutativity for Neveu-Schwarz boundary conditions}}}
\author{
{\bf {\normalsize Chandrasekhar Chatterjee}$^{a}$\thanks{chandra@bose.res.in}},\,
 {\bf {\normalsize Sunandan Gan{g}opadhyay}$^{a}$\thanks{sunandan@bose.res.in}} \\
 {\bf {\normalsize Arindam Ghosh Hazra}$^{b}$\thanks{arindamg@bose.res.in}},\,\,{\bf {\normalsize Saurav Samanta}$^{a}$\thanks{saurav@bose.res.in}}
\\
$^{a}$ {\normalsize S.~N.~Bose National Centre for Basic Sciences,}\\{\normalsize JD Block, Sector III, Salt Lake, Kolkata-700098, India}\\[0.3cm]
$^{b}$ {\normalsize Department of Mathematics, Sundarban Mahavidyalaya}\\{\normalsize Kakdwip, South 
24 Parganas, West-Bengal, India}\\[0.3cm]
}
\date{}

\maketitle

\begin{abstract}
The appearance of non(anti)commutativity in superstring theory, 
satisfying the Neveu-Schwarz boundary conditions is discussed in this paper. 
Both an open free superstring and also one moving in a 
background antisymmetric tensor field are analyzed 
to illustrate the point that string non(anti)commutativity 
is a consequence of the nontrivial boundary conditions. 
The method used here is quite different 
from several other approaches where boundary conditions 
were treated as constraints. 
An interesting observation of this study is that, one 
requires that the bosonic sector satisfies 
Dirichlet boundary conditions at one end 
and Neumann at the other 
in the case of the bosonic variables $X^{\mu}$ being antiperiodic.
The non(anti)commutative structures derived in this paper 
also leads to the closure of the 
super constraint algebra which is essential for the internal consistency 
of our analysis. 

\\[0.3cm]

\end{abstract}

\section{Introduction}
It is now a well established fact theoretically 
that at fundamental scale, 
space time becomes noncommutative (NC) in nature. This can be understood 
from a fundamental theory namely string theory where in the 
presence of a background antisymmetric $\cal{B}_{\mu\nu}$ field, the 
string end points 
become NC \cite{sw,dn} and therefore the embedding 
$D$-brane coordinates also become NC. 
A detailed discussion of this problem in the context of 
bosonic string has been done in \cite{chu,som,zab}. 
On the other hand noncommutativity for both the 
bosonic and the fermionic string theories has 
been explored in \cite{sheikh,ard, godinho,ohta}. 
In \cite{chu,sheikh,ard,godinho} the Neumann and 
Dirichlet boundary conditions (BC) which get mixed due to the presence 
of non vanishing $\cal{B}$ field were 
considered as Hamiltonian constraints. A different approach 
was followed in \cite{zab} 
where the constraints appear in an algebraic manner 
within the framework of Hamiltonian formulation. 
In \cite{sheikh} noncommutativity appears 
not only at the boundary but also in the bulk, 
where as in \cite{kim} noncommutativity 
was shown to be exist only at the boundary. 
Andrade et. al.\cite{and} took the 
BC(s) as second class constraint and project the 
original coordinates on the constraint surface 
to recover a set of unconstrained string coordinates. 
Very recently the Faddeev-Jackiw symplectic formalism and a conformal
field theoretic approach have been used to obtain the NC structure in 
\cite{jing2,jing1,sg, sg1}.

\noindent However, there is an alternative method first 
initiated by Hanson, Regge and 
Teitelboim \cite{hrt} for the free Nambu-Goto string
and later extended in \cite{rb} for the Polyakov string, 
where instead of taking the BC(s) as constraints 
it is shown that the NC structure 
appears as a natural modification of the 
Poisson bracket (PB) to make it compatible with the BC(s). 
Previously in a couple of papers \cite{agh,as}, 
two of the authors used this approach to investigate this problem. 
Surprisingly a common point of all these studies in the 
superstring theory is that all the literatures 
are solely confined for the Ramond (R) BC(s) only. 
But as is well known in the case of fermionic string there 
is a choice between R BC(s) and Neveu Schwarz (NS) BC(s). 
This second type of BC is less studied in the research area. 
Here in this paper, we extend the above mentioned 
methodology to the superstring satisfying the NS BC(s). 
A nontrivial result we  have found from the whole 
analysis is that contrary to the R case, 
bosonic sector of the superstring satisfies 
Dirichlet BC at one end and Neumann BC at the other end
provided the bosonic variable $X^{\mu}$ is allowed to be antiperiodic. 
This observation is completely 
new and has not been discussed elsewhere.
Further, the symplectic structure of the bosonic
sector also keeps the 
superconstraint algebra involutive.
The bracket structures have also been computed
using the mode expansions of the bosonic and the fermionic 
coordinates.

\noindent The organization of the paper is as follows: 
In section 2, the R-Neveu Schwarz (RNS)
superstring action in the conformal gauge
is briefly discussed to fix the notations.
The section is then subdivided into two parts.
In the first subsection, the BC(s) and the mode expansions 
of the fermionic sector
of the superstring is given and the nonanticommutativity
of the theory is revealed in the
conventional Hamiltonian framework. In the next subsection, the 
PB structure and the BC(s) of the bosonic sector is discussed.
In section 3, we compute the super constraint algebra
with the modified symplectic structure obtained in the previous section. The results obtained in section 2 is further confirmed in section 4 by the mode expansion method. This consistency check is performed separately for the bosonic and the fermionic sector.
Section 5 discusses the non(anti)commutativity in the interacting
superstring theory in the RNS formulation. Finally section 6 is for conclusions.


\section {RNS free superstring}
In the first part of this section we briefly 
mention the canonical algebra of the 
basic fields of a free open superstring. 
Later we shall show how these algebraic structures 
get modified as a result of the boundary conditions of the theory. 
The action we take 
for our analysis is given by \cite{green}\footnote{We follow the conventions  
$\rho^0 \,=\, \sigma^2 \, = \, 
\pmatrix{0&-i\cr i&0\cr}\,\,\,,\,\,\,
\rho^1 \,=\, i\sigma^1 \, = \, \pmatrix{0&i\cr i&0\cr}\ $ and take the 
induced world-sheet metric and target space-time metric as 
$\eta^{ab} = \{-, +\}$,  
 $\eta^{\mu\nu} =\{-, +, +, ...., +\}$ respectively.} 
\begin{eqnarray}
\label{1}
S &=& -\frac{1}{2}
\int_{\Sigma} d^2\sigma \Big(
\eta_{\mu \nu } \partial_a X^\mu \partial^a X^\nu 
\,- \,  i {\overline \psi}^\mu \rho^a \partial_ a \psi_\mu \Big).
\end{eqnarray}
The bosonic and the fermionic part of the above action 
can be separated out as
\begin{eqnarray}
S &=& S_B + S_F
\end{eqnarray}
where,
\begin{eqnarray}
\label{1aa}
S_B = -\frac{1}{2}\int_{\Sigma} d^2\sigma \eta_{\mu \nu }
\partial_a X^\mu \partial^a X^\nu \  \ {\textrm{and}} \  
\ S_F = \frac{1}{2}\int_{\Sigma} d^2\sigma
i {\overline \psi}^\mu \rho^a \partial_ a \psi_\mu.  
\end{eqnarray}
The components of the Majorana spinor $\psi$ are denoted as $\psi_{\pm}$
\begin{equation}
\psi^\mu \,=\, \pmatrix{\psi^\mu_{-}\cr \psi^\mu_{+}\cr}\, .
\label{3}
\end{equation}
\noindent The Dirac antibracket of the first order action 
$S_F$ is easily read off 
\begin{eqnarray}
\{ \psi^\mu_{+} (\sigma) \,,\,\psi^\nu_{+}  (\sigma^\prime) \}_{D.B} &=& 
\{ \psi^\mu_{-} (\sigma) \,,\,\psi^\nu_{-} (\sigma^\prime ) \}_{D.B}
\,\,=\,\,
 - i \eta^{\mu\nu} \delta (\sigma - \sigma^\prime )\,
\nonumber\\
\{ \psi^\mu_{+}(\sigma)\,,\,\psi^\nu_{-}(\sigma^\prime) \}_{D.B} &=& 0\,\,.
\label{5}
\end{eqnarray}
On the other hand the action $S_B$ gives the 
following brackets among the bosonic variables
\begin{eqnarray}
&&\{X^\mu(\sigma) , \Pi^\nu(\sigma^{\prime})\} = 
\eta^{\mu \nu}\delta(\sigma - \sigma^{\prime})\nonumber\\
&&\{X^\mu(\sigma) , X^\nu(\sigma^{\prime})\} =0=\{\Pi^\mu(\sigma) , 
\Pi^\nu(\sigma^{\prime})\}
\label{18a}
\end{eqnarray}
where $\Pi_\mu$ is the canonically conjugate momentum
to $X^\mu$, defined in the usual way. Eqs. (\ref{5}) and (\ref{18a})
defines the preliminary symplectic structure of the theory. 
We shall now discuss the effects of BC(s) on these symplectic algebra for the fermionic and the bosonic sectors
separately.
\subsection{Fermionic sector}
Varying the fermionic part of the action (\ref{1aa})
\begin{eqnarray}
\label{18}
\delta S_F = i \int_{\Sigma} d^2\sigma 
\left[ \delta \bar{\psi}_{\mu}\rho^a \, \partial_a 
\psi^{\mu} \, - \partial_{\sigma}
\left(\psi^\mu_{-}\, \delta \psi_{\mu -} - 
\psi^\mu_{+}\, \delta \psi_{\mu +}\right)\right]
\end{eqnarray}
we obtain the Euler-Lagrange equation for the fermionic field 
\begin{eqnarray}
\label{10}
i \rho^a \partial_a \psi^{\mu} =0.
\end{eqnarray}
together with the following BC(s): 
\begin{eqnarray}
\psi^{\mu}_{+}(0 , \tau) &=& \psi^{\mu}_{-}(0 , \tau) \nonumber \\
\psi^{\mu}_{+}(\pi , \tau) &=& \lambda \psi^{\mu}_{-}(\pi , \tau)
\label{12}
\end{eqnarray}
where $\lambda = \pm 1$ corresponds to the R BC(s) and the
NS BC(s), respectively.
The investigation involving the R BC(s) has been made in  
\cite{jing1, sg1, agh}. 
In this paper, we shall work with the NS BC(s) 
which we write in the following manner
\begin{eqnarray}
\label{6aaa}
(\psi^{\mu}_{+}(\sigma , \tau) - \psi^{\mu}_{-}(\sigma , \tau))|_{\sigma=0}=0
\end{eqnarray}
\begin{eqnarray}
\label{13aaa}
(\psi^{\mu}_{+}(\sigma , \tau) + \psi^{\mu}_{-}(\sigma , \tau))|_{\sigma=\pi}=0.
\end{eqnarray}
Now the mode expansion of the components of Majorana fermion, 
satisfying the above set of BC(s) is given by \cite{green}:
\begin{eqnarray}
\psi^{\mu}_{-}(\sigma , \tau) &=& \frac{1}{\sqrt{2}}\sum_{n\in Z + 
\frac{1}{2}} d^{\mu}_{n}
e^{-i\, n(\tau - \sigma )} \nonumber \\ 
\psi^{\mu}_{+}(\sigma , \tau) &=& \frac{1}{\sqrt{2}}\sum_{n\in Z + 
\frac{1}{2}} d^{\mu}_{n}
e^{-i\,n(\tau + \sigma )}.
\label{7}
\end{eqnarray}
From the above mode expansions it follows automatically that
\begin{eqnarray}
\psi^{\mu}_{-}(-\sigma , \tau) \,=\, \psi^{\mu}_{+}(\sigma , \tau)\,.
\label{8}
\end{eqnarray}
Furthermore, making use of eq. (\ref{12}), we obtain
\begin{eqnarray}
\psi^{\mu}_{\pm}(\sigma = -  \pi , \tau)
&=& - \psi^{\mu}_{\pm}(\sigma=\pi , \tau) \nonumber\\
\psi^{\mu}_{\pm}(\sigma = -  2\pi , \tau)
&=&  \psi^{\mu}_{\pm}(\sigma = 2\pi , \tau)
\label{9}
\end{eqnarray}
in the NS-sector. Hence $\psi^{\mu}_{\pm}(\sigma , \tau)$ 
is an antiperiodic function of antiperiodicity $2\pi$ which naturally implies that it is a periodic function of periodicity $4\pi$.
\noindent We now essentially follow the methodology of \cite{agh}
for the present case. First, we introduce the antiperiodic delta function
$\delta_{(a)P}(x)$ of antiperiodicity $2\pi$ and periodicity $4\pi$
\begin{eqnarray}
\delta_{(a)P}(x)= - \delta_{(a)P}(x + 2\pi)=\frac{1}{4\pi}
\sum_{n\in Z + 
\frac{1}{2}}e^{i\,n x}
\label{ap}
\end{eqnarray}
which satisfies the defining property of a periodic $\delta$-function i.e.
\begin{eqnarray}
\int_{-2\pi}^{2\pi}dx^{\prime}\delta_{(a)P}(x^{\prime}-x)f(x^{\prime})=f(x)
\label{40}
\end{eqnarray}
where $f(x)$ is an arbitrary periodic function 
with periodicity $4\pi$. 
Using this we write the following 
expression for $\psi^{\mu}_{-}$ and $\psi^{\mu}_{+}$ 
in the physical interval $[0,\pi]$ of the string
\begin{eqnarray}
2\int_{0}^{\pi}d\sigma^{\prime}\left[\delta_{(a)P}(\sigma^{\prime} +
\sigma)\psi^{\mu}_{+}(\sigma^{\prime}) + 
\delta_{(a)P}(\sigma^{\prime}-\sigma)\psi^{\mu}_{-}(\sigma^{\prime})
\right] &=& \psi^{\mu}_{-}(\sigma)\,
\label{41} \\
2\int_{0}^{\pi}d\sigma^{\prime}\left[\delta_{(a)P}(\sigma^{\prime} 
+ \sigma)\psi^{\mu}_{-}(\sigma^{\prime}) + 
\delta_{(a)P}(\sigma^{\prime}-\sigma)\psi^{\mu}_{+}(\sigma^{\prime})
\right] &=& \psi^{\mu}_{+}(\sigma)\,.
\label{42}
\end{eqnarray}
We define a matrix $\Lambda_{AB}(\sigma, \sigma^{\prime})$
\begin{equation}
\Lambda_{AB}(\sigma, \sigma^{\prime})\,=\, 
\pmatrix{\delta_{(a)P}(\sigma^{\prime}-\sigma)&
\delta_{(a)P}(\sigma^{\prime}+\sigma)\cr
\delta_{(a)P}(\sigma^{\prime}+\sigma)&
\delta_{(a)P}(\sigma^{\prime}-\sigma)\cr}\,
\label{44}
\end{equation}
to write the equations (\ref{41}) and (\ref{42}) in a compact form
\begin{eqnarray}
2\int_{0}^{\pi}d\sigma^{\prime}\Lambda_{AB}(\sigma, \sigma^{\prime}) 
\psi^{\mu}_{B}(\sigma^{\prime})=\psi^{\mu}_{A}(\sigma)\quad;\quad (A, \ B=-,+).
\label{43}
\end{eqnarray}
From the above equation $\Lambda$ can be interpreted as a matrix valued ``delta function" which acts on the two component Majorana spinor. Instead of (\ref{5}) we therefore propose the following antibrackets in the fermionic sector
\begin{equation}
\{\psi^{\mu}_{A}(\sigma), \psi^{\nu}_{B}(\sigma^{\prime})\}
=-2i\eta^{\mu\nu}\Lambda_{AB}(\sigma, \sigma^{\prime}).
\label{45}
\end{equation}
Making use of eq. (\ref{44}) we write this in its component form
\begin{eqnarray}
\{\psi^{\mu}_{+}(\sigma), \psi^{\nu}_{+}(\sigma^{\prime})\}
&=&\{\psi^{\mu}_{-}(\sigma), \psi^{\nu}_{-}(\sigma^{\prime})\}
=-2i\eta^{\mu\nu}\delta_{(a)P}(\sigma - \sigma^{\prime})\,\nonumber\\
\{\psi^{\mu}_{-}(\sigma), \psi^{\nu}_{+}(\sigma^{\prime})\}
&=&-2i\eta^{\mu\nu}\delta_{(a)P}(\sigma + \sigma^{\prime})\,.
\label{46}
\end{eqnarray}
Remarkably the above set of antibracket algebra is now completely
consistent with the BC(s). To see this explicitly,
we compute the anticommutator of $\psi_{+}^{\nu}(\sigma^{\prime})$ with 
(\ref{6aaa}) and (\ref{13aaa}), the left hand side of which gives:
\begin{eqnarray}
-2i\left(\delta_{(a)P}(\sigma - \sigma^{\prime})-
\delta_{(a)P}(\sigma + \sigma^{\prime})\right)|_{\sigma = 0}&=&
- 2i \Delta_{-(a)}\left(\sigma , \sigma^{\prime}\right)|_{\sigma = 0}
\nonumber\\
&=& \frac{i}{\pi}\sum_{n \in Z + \frac{1}{2}}{\textrm{sin}}(n\sigma) 
\ {\textrm{sin}}
(n \sigma')|_{\sigma = 0}=0
\label{2237}\\
-2i\left(\delta_{(a)P}(\sigma - \sigma^{\prime})+
\delta_{(a)P}(\sigma + \sigma^{\prime})\right)|_{\sigma = \pi}&=&
- 2i \Delta_{+(a)}\left(\sigma , \sigma^{\prime}\right)|_{\sigma = \pi}
\nonumber\\
&=& -\frac{i}{\pi}\sum_{n \in Z + \frac{1}{2}}{\textrm{cos}}(n \sigma) 
\ {\textrm{cos}}
(n \sigma')|_{\sigma = \pi}= 0
\label{47}
\end{eqnarray}
where the form of the antiperiodic delta function (\ref{ap}) has been used.
This completes the analysis of the fermionic algebra for the NS BC(s). 
In the next section we shall use these 
relations (\ref{46}) to compute the super constraint algebra.
\subsection{Bosonic sector}
Let us now study the bosonic sector of the superstring action (\ref{1}). 
Varying the bosonic part of the action (\ref{1}), 
we obtain the equation of motion for the bosonic
field
\begin{eqnarray}
(\partial_{\sigma}^2-\partial_{\tau}^2)X^{\mu} = 0
\label{eqb}
\end{eqnarray}
together with Dirichlet and Neumann BC(s)
\begin{eqnarray}
\delta X^{\mu}|_{\sigma=0,\pi} = 0 \nonumber \\
X'^{\mu}|_{\sigma=0,\pi}=0.
\label{bcb}
\end{eqnarray}
Now there are two cases depending on the periodicity of the bosonic variable
$X^{\mu}$. Usually, one is interested in theories with maximum
Poincar\'{e} invariance and hence $X^{\mu}$ must be periodic (with a 
periodicity of $2\pi$). This case has already been discussed in \cite{agh}.
On the other hand antiperiodicity of $X^{\mu}$ is interesting because
one encounters it for twisted strings on an orbifold \cite{polc}. 
In this paper we
shall discuss this case in details.

\noindent We let the bosonic string coordinates 
$X^{\mu} (\sigma)$ to have a periodicity of 
$4 \pi$ (antiperiodicity of $2\pi$)\footnote{Note that this is 
also in accord with the fermionic sector.}:
\begin{eqnarray}
X^{\mu}(\sigma+4\pi)=X^{\mu}(\sigma).
\label{periodic}
\end{eqnarray}
Hence the integral (\ref{40}) once again holds for the bosonic coordinate 
$X^{\mu}(\sigma)$. Restricting to the case of even(odd)
functions $X^{\mu}_{\pm}(- \sigma) = \pm X^{\mu}_{\pm}(\sigma)$,
it can be easily seen that (\ref{40}) reduces to:
\begin{eqnarray}
2\int^{\pi}_{0} d\sigma^{\prime}\, \Delta_{\pm(a)}(\sigma,\sigma')\,
X^{\mu}_{\pm}(\sigma^{\prime}) \,=\, X^{\mu}_{\pm}(\sigma) 
\label{apb}
\end{eqnarray}
where $\Delta_{\pm(a)}$ were defined in the eqs. 
(\ref{2237}) and (\ref{47}). Wetherefore propose 
the following equal time PB:
\begin{eqnarray}
\left\{X^{\mu}(\tau , \sigma) ,  \Pi_{\nu}(\tau , \sigma^{\prime})\right\}
= 2\,\delta^{\mu}_{\nu}\, \Delta_{\pm(a)}(\sigma, \sigma^{\prime}).
\label{etpb}
\end{eqnarray}
It is now easy to observe that for $\Delta_{+ (a)}(\sigma,\sigma')$
to appear in the above
PB the end points must satisfy following BC(s)
\begin{eqnarray}
&& X^{' \mu}(0) = 0 \nonumber\\
&& X^{\mu}(\pi) = 0
\label{apbcb}
\end{eqnarray}
and for $\Delta_{- (a)}(\sigma,\sigma')$, the appropriate BC(s) 
that the end points must satisfy read
\begin{eqnarray}
&& X^{\mu}(0) = 0 \nonumber\\
&& X^{' \mu}(\pi) = 0.
\label{apbcb1}
\end{eqnarray}
We shall find in the next section that the symplectic structure
of the bosonic sector also plays a crucial role in the
closure of the super constraint algebra.

\section {Super constraint algebra }
In this section we shall compute the algebra of the super-Virasoro 
constraints using the modified symplectic structures derived in section 2.\\
\noindent The complete set of super constraints 
are given by \cite{ agh,green}:
\begin{eqnarray}
\chi_1(\sigma) &=& \Phi_1(\sigma) + \lambda_1(\sigma)
\approx 0 \nonumber \\ 
\chi_2(\sigma) &=& 
\Phi_2(\sigma) + \lambda_2(\sigma) \approx 0
\label{chi}
\end{eqnarray}
where,
\begin{eqnarray}
\Phi_1(\sigma) &=& 
\left(\Pi^2(\sigma) + (\partial_{\sigma}X(\sigma))^2\right) \nonumber \\ 
\Phi_2(\sigma) &=& 
\left(\Pi(\sigma)\partial_{\sigma}X(\sigma)\right) \nonumber \\
\lambda_1(\sigma)&=&  
- i \bar{\psi^{\mu}}(\sigma)\rho_{1} \partial_{\sigma} \psi_{\mu}(\sigma)
=  - i \left(\psi^{\mu}_{-}(\sigma) 
\partial_{\sigma} \psi_{\mu -}(\sigma) - \psi^{\mu}_{+}(\sigma) 
\partial_{\sigma} \psi_{\mu +}(\sigma)\right) \nonumber \\ 
\lambda_2(\sigma)&=& 
- \frac{i}{2}  \bar{\psi^{\mu}}(\sigma)\rho_{0} 
\partial_{\sigma} \psi_{\mu}(\sigma)  
= \frac{i}{2}\left(\psi^{\mu}_{-}(\sigma)\partial_{\sigma} \psi_{\mu -}(\sigma) + \psi^{\mu}_{+}(\sigma)\partial_{\sigma} \psi_{\mu +}(\sigma)\right)
\label{lambda}
\end{eqnarray}
and
using the basic algebra of fermionic and bosonic variables 
(\ref{46}, \ref{etpb}), we get the following algebra for 
super-Virasoro constraints:
\begin{eqnarray}
\{\chi_1(\sigma) , \chi_1(\sigma^{\prime})\} &=& 8 \left(
\chi_2(\sigma)\partial_{\sigma}\Delta_{+ (a)}
\left(\sigma , \sigma^{\prime}\right) + \chi_2(\sigma^{\prime})
\partial_{\sigma}\Delta_{- (a)}\left(\sigma , \sigma^{\prime}\right)\right)\,
\nonumber \\ 
\{\chi_2(\sigma) , \chi_2(\sigma^{\prime})\} &=&  2\left(
\chi_2(\sigma^{\prime})\partial_{\sigma}\Delta_{+ (a)}
\left(\sigma , \sigma^{\prime}\right) + \chi_2(\sigma)
\partial_{\sigma}\Delta_{- (a)}\left(\sigma , \sigma^{\prime}\right)\right) \,
\nonumber \\ 
\{\chi_2(\sigma) , \chi_1(\sigma^{\prime})\} &=&  2\left(
\chi_1(\sigma) + \chi_1(\sigma^{\prime})\right)
\partial_{\sigma}\Delta_{+ (a)}\left(\sigma , \sigma^{\prime}\right)\,.
\label{chi-chi}
\end{eqnarray}
Apart from a numerical factor the above algebra 
has the same structure
as in \cite{agh} with the only difference that $\delta_{P}(\sigma)$
occurring in \cite{agh} has been replaced by 
$\delta_{(a)P}(\sigma)$. Similarly one can show that the 
algebra of super currents
\begin{eqnarray}
&&\tilde{J_1}(\sigma)=2J_{01}(\sigma)=\psi_-^{\mu}(\sigma)\Pi_{\mu}(\sigma)-\psi_-^{\mu}(\sigma)\partial_{\sigma}X_{\mu}\nonumber\\
&&\tilde{J_2}(\sigma)=2J_{02}(\sigma)=\psi_+^{\mu}(\sigma)\Pi_{\mu}(\sigma)+\psi_+^{\mu}(\sigma)\partial_{\sigma}X_{\mu}
\label{Jtilde}
\end{eqnarray}
among themselves and also with the super constraints (\ref{chi}) close. It is also interesting to note that both 
$\Delta_{+(a)}$ and $\Delta_{-(a)}$ appearing in the
PB of the bosonic variables (\ref{etpb}) gives the same
constraint algebra (\ref{chi-chi}). Furthermore, the closure of the algebra also 
indicates the internal consistency of our analysis.
\section{Mode expansions and symplectic algebra}
In this section, we shall derive the fermionic algebra (\ref{46}) and the bosonic algebra (\ref{etpb}) from a mode expansion of the constituting fields. 
To do that we consider the mode expansions of the fermionic field 
(\ref{7}). Here $d^{\mu}_{n}$ are Fourier modes and
they satisfy the algebra
\begin{eqnarray}
\left\{d^{\mu}_{m}, d^\nu_n\right\} = 
-\frac{i}{\pi}\, \eta^{\mu \nu}\, \delta_{m+n,0}.
\label{s0}
\end{eqnarray}
This algebra can be obtained just by  following the procedure
of \cite{jing2}, in which they have computed the anti brackets among Fourier components of fermionic sector of superstrings
(R sector) using Faddeev-Jackiw symplectic formalism 
\cite{fad}. This relation (\ref{s0}) between $d$'s can also be worked 
out from the contour argument \cite{sg, polc} and the operator
product expansion.
\noindent
The antibracket relations between $\psi^{\mu}_{A}(\s),
\psi^{\nu}_{B}(\s^{\prime})$ are then obtained by using
(\ref{7}) and (\ref{s0})
\begin{eqnarray}
\left\{\psi^{\mu}_{-}(\s), \psi^{\nu}_{+}(\s^{\prime})\right\} \, &=& \, \frac{1}{2} \sum_{r, s\in{Z + \frac{1}{2}}}
e^{- i r (\tau - \s)}\, e^{- i s (\tau +\s)}
\left\{d^{\mu}_{r} , d^{\nu}_{s}\right\} \\ \no
&=& - \frac{i}{2\pi}\, \eta^{\mu \nu}\sum_{r\in{Z + \frac{1}{2}}}e^{- i r (\tau - \s)}\, e^{i r(\tau +\s)}\\ \no
&=& -2 i \eta^{\mu \nu} \delta_{(a)P}(\s + \s^\prime).
\label{s1}
\end{eqnarray}
Proceeding exactly in the similar manner one can get back the other 
anti-brackets of (\ref{46}).

\noindent
In order to study the bosonic sector, we first need the expressions
 of the mode expansion for the two different types of BC(s) (\ref{apbcb})
 and (\ref{apbcb1}). \\
\noindent
For the first case (BC (\ref{apbcb})) it is given by:
\begin{eqnarray}
X^{\mu}(\tau, \s) = \sum_{n\in{Z + \frac{1}{2}}}
\frac{\alpha^{\mu}_{n}}{n}\, e^{in\tau}\, \mathrm{sin}\,n\s
\label{s2}
\end{eqnarray}
and for the other case (BC (\ref{apbcb1})) the mode expansion is
\begin{eqnarray}
X^{\mu}(\tau, \s) = \sum_{n\in{Z + \frac{1}{2}}}
\frac{\alpha^{\mu}_{n}}{n}\, e^{in\tau}\, \mathrm{cos}\,n\s.
\label{s3}
\end{eqnarray}
The canonical momenta corresponding to (\ref{s2}) and (\ref{s3}) are given by
\begin{eqnarray}
\Pi_{\mu}(\tau, \s) &=& \eta_{\mu \nu} \p_{\tau}X^{\nu}(\tau,
\s)  \no\\
&=& i\eta_{\mu \nu}\sum_{n\in{Z + \frac{1}{2}}}
\alpha^{\nu}_{n}\, e^{in\tau}\, \mathrm{sin}\,n\s\\
&=& i\eta_{\mu \nu}\sum_{n\in{Z + \frac{1}{2}}}
\alpha^{\nu}_{n}\, e^{in\tau}\, \mathrm{cos}\,n\s.
\label{s5}
\end{eqnarray}
Here also the algebra between the modes can be computed by following the methodology of \cite{jing1, fad}:
\begin{eqnarray}
\left\{\alpha^{\mu}_{m}, \alpha^\nu_n\right\} = 
- \frac{i}{\pi}\, \eta^{\mu \nu}\, m\delta_{m+n,0}.
\label{s4}
\end{eqnarray}
Using (\ref{s4}) we obtain the same equal time PB given in (\ref{etpb}).
\section{The interacting theory}
After finishing the analysis for the free theory, 
we shall now study the interacting case where a 
superstring moves in the presence of a 
constant antisymmetric tensor field ${\cal B}_{\mu \nu}$. 
The action given by \cite{agh, haggi, NL}:
\begin{eqnarray}
S &=&  \frac{- 1}{2} \int_{\Sigma} d\tau
 d{\s}\,\Big[ \, \partial_a X^\mu \partial^a X_\mu 
\,+\, \epsilon^{ab} {\cal B}_{\mu\nu} \partial_a X^\mu \partial_b X^\nu
\nonumber\\
 & & + i \psi_{\mu -} E^{\nu \mu} \partial_{+} \psi_{\nu -} +
i \psi_{\mu +} E^{\nu \mu} \partial_{-} \psi_{\nu +} \,\Big]
\label{1.1}
\end{eqnarray}
\noindent where, $\partial_{+} =  \partial_{\tau} + \partial_{\s},\; \;
\partial_{-} =  \partial_{\tau} - \partial_{\s} $
and $ E^{\mu\nu} \,=\, \eta^{\mu\nu} \, + {\cal B}^{\mu\nu}\,$.
\noindent Now since the bosonic and fermionic sectors decouple, we 
can study them separately.
\noindent
Here we concentrate on the fermionic sector. 
The variation of the fermionic part of the action (\r{1.1})
gives the classical equations of motion: 
\begin{equation}
\partial_{+} \psi_{\nu -} \,=\, 0\quad,\quad
\partial_{-} \psi_{\nu +} \,=\, 0
\label{1.2}
\end{equation}
and a boundary term that yields the following NS BC(s)\footnote{
The boundary term also leads to R BC(s). Detailed investigations
involving R BC(s)
has already been carried out in \cite{agh, haggi, NL}.}:
\begin{eqnarray}
 E_{\nu\mu}\,  \psi^\nu_{+} (0,\tau)\, &=&\,
  E_{\mu\nu} \, \psi^\nu_{-} (0,\tau)\, \nonumber\\
\label{4}
E_{\nu\mu}\,  \psi^\nu_{+} (\pi,\tau ) \, &=& \, -
E_{\mu\nu} \, \psi^\nu_{-} (\pi, \tau )
\label{1.3}
\end{eqnarray}
\noindent at the endpoints $\s \,=\,0$ and $\s = \pi\,$ of the string.

\noindent As in the free case, the above non-trivial BC(s) 
leads to a modification 
in the symplectic structure (\ref{5}). The $\{\psi^\mu_{(\pm)} (\sigma,\tau) , 
\psi^\nu_{\pm} (\sigma^{\prime},\tau)\}$ is the same as 
(\ref{46}). In the case of mixed
bracket, we make the following ansatz:
\begin{eqnarray}
 \{\psi^\mu_{+} (\sigma,\tau) , 
\psi^\nu_{-} (\sigma^{\prime},\tau)\}\, = \, 
C^{\mu\nu}\delta_{(a)P}\left(\sigma + \sigma^{\prime}\right)\,.
\label{57}
\end{eqnarray}
Brackets $\psi^\gamma_{-} (\sigma^{\prime})$ with the BC(s) (\ref{1.3}) one obtains
\begin{eqnarray}
 E_{\nu\mu}\,  C^{\nu \gamma}
\, = \, -2i \, E_{\mu\gamma}
\label{58}
\end{eqnarray}
which on solving gives
\begin{eqnarray}
 C^{\mu \nu}\, = \, -2i \, 
\left[\left(1 - {\cal B}^2\right)^{-1}\right]^{\mu \rho}\, 
 E_{\rho \gamma}\, E^{\gamma \nu}.
\label{58a}
\end{eqnarray}
Above solution is written in a matrix notation as, 
\begin{eqnarray}
 C \, = \, -2i \, 
\left[\left(1 - {\cal B}^2\right)^{-1}\left(1 + {\cal B}\right)^{2}\right]
\label{58y}
\end{eqnarray}
where $C = \{C^{\mu \nu}\}$. Thus we get the modified mixed bracket in the form
\begin{eqnarray}
\{\psi^\mu_{+} (\sigma,\tau) , 
\psi^\nu_{-} (\sigma^{\prime},\tau)\}\, = \, 
-2i\, \left[\left(1 - {\cal B}^2\right)^{-1}\right]^{\mu \rho}
\,  E_{\rho \gamma}\, E^{\gamma \nu}
\delta_{(a)P}\left(\sigma + \sigma^{\prime}\right)\,.
\label{60}
\end{eqnarray}
If we take the limit ${\cal B}_{\mu \nu} \rightarrow 0$ in the above equation we get back the last relation of (\ref{46}).
\section{Conclusions}
In string theory the modification of Poisson algebra 
is a consequence of the nontrivial BC(s). 
In this paper, we have studied this problem for 
an open superstring satisfying the NS BC(s). 
Following the approach of \cite{agh}, 
here also the domain of the string length is extended from $[0, \pi]$ to $[-\pi, \pi]$ to get the antiperiodic BC(s). This construction enables us to get the $2 \times 2$ matrix valued $\delta$ function in the algebra of the fermionic sector. Apart from a numerical factor the fermionic algebra is identical to the result obtained in \cite{agh}. In this sense this paper is an extension of\cite{agh}, where only R BC(s) was used.

\noindent
However for the bosonic part of the superstring the result of 
present paper is drastically different \cite{agh}. 
We stress that the symplectic algebra of the bosonic 
variables, in this paper contains both 
$\Delta_{+(a)}(\sigma , \sigma^{\prime})$ and 
$\Delta_{-(a)}(\sigma , \sigma^{\prime})$ which 
is completely different from the R case where only 
$\Delta_{+}(\sigma , \sigma^{\prime})$ was present. 
Interestingly this symplectic structure containing 
both $\Delta_{\pm(a)}(\sigma , \sigma^{\prime})$, 
keeps the superconstraint algebra closed provided 
one imposes Neumann BC(s) at one end and Dirichlet 
BC(s) at the other end of the string in the bosonic sector. 
This observation is completely new and has not been 
noticed before in the literature.
\noindent
Finally to complete the analysis, we have calculated 
the non(anti)commutative structures for the interacting 
case by employing the same procedure. 
As one expects, without the background field term the 
interacting results take the limiting value of the free case.

\end{document}